\documentclass[%
 reprint,
 amsmath,amssymb,
 aps,
]{revtex4-1}

\usepackage{graphicx}
\usepackage{dcolumn}
\usepackage{bm}
\usepackage{color}

\newcommand{\R}{^R}
\newcommand{\tI}{t_a}
\newcommand{\tF}{t_b}

\newcommand{\fun}{{\cal F}}
\newcommand{\syst}{{\bf S}}

\newcommand{\xx}{x}
\newcommand{\xxI}{\xx_a}
\newcommand{\xxF}{\xx_b}

\newcommand{\yw}{\alpha}

\newcommand{\ywF}{\yw_b}

\newcommand{\sel}{{\cal J}}

\newcommand{\Gh}{\Gamma}
\newcommand{\GA}{\Gh_\syst}

\newcommand{\Gf}{G_b}
\newcommand{\Gs}{G_a}

\newcommand{\OS}{{\cal O}}

\newcommand{\info}{{\cal I}}
\newcommand{\infA}{\info_a}
\newcommand{\infB}{\info_b}

\newcommand{\pr}{\pi}
\newcommand{\pf}{\overline{\pi}}

\newcommand{\pin}{P_{a\to b}}
\newcommand{\pout}{P_{b\to a}}
\newcommand{\rec}{{M}}
\newcommand{\e}{e}

\begin{document}

\preprint{APS/123-QED}

\title{Information break of time symmetry in the macroscopic limit}

\author{Miroslav Hole\v{c}ek}
 \email{holecek@rek.zcu.cz}
\affiliation{%
 New Technologies Research Center, University of West Bohemia, Plze\v{n} 301 00,  Czech Republic
}%

\date{\today}

\begin{abstract}

The evident contrast between the time symmetry of fundamental microscopic laws  and the time asymmetry of macroscopic processes is a challenging  physical problem. The observation of unitary evolution of a general physical system by an idealized observer whose recording and processing information is time neutral is studied in a full information context.  In the macroscopic limit, a contradiction between observations and predictions  done at different times appears. 
It implies that the observer cannot read records performed at times when the entropy of the system is larger.

\end{abstract}

\maketitle

\begin{figure}[h]
\includegraphics[width=0.4\textwidth]{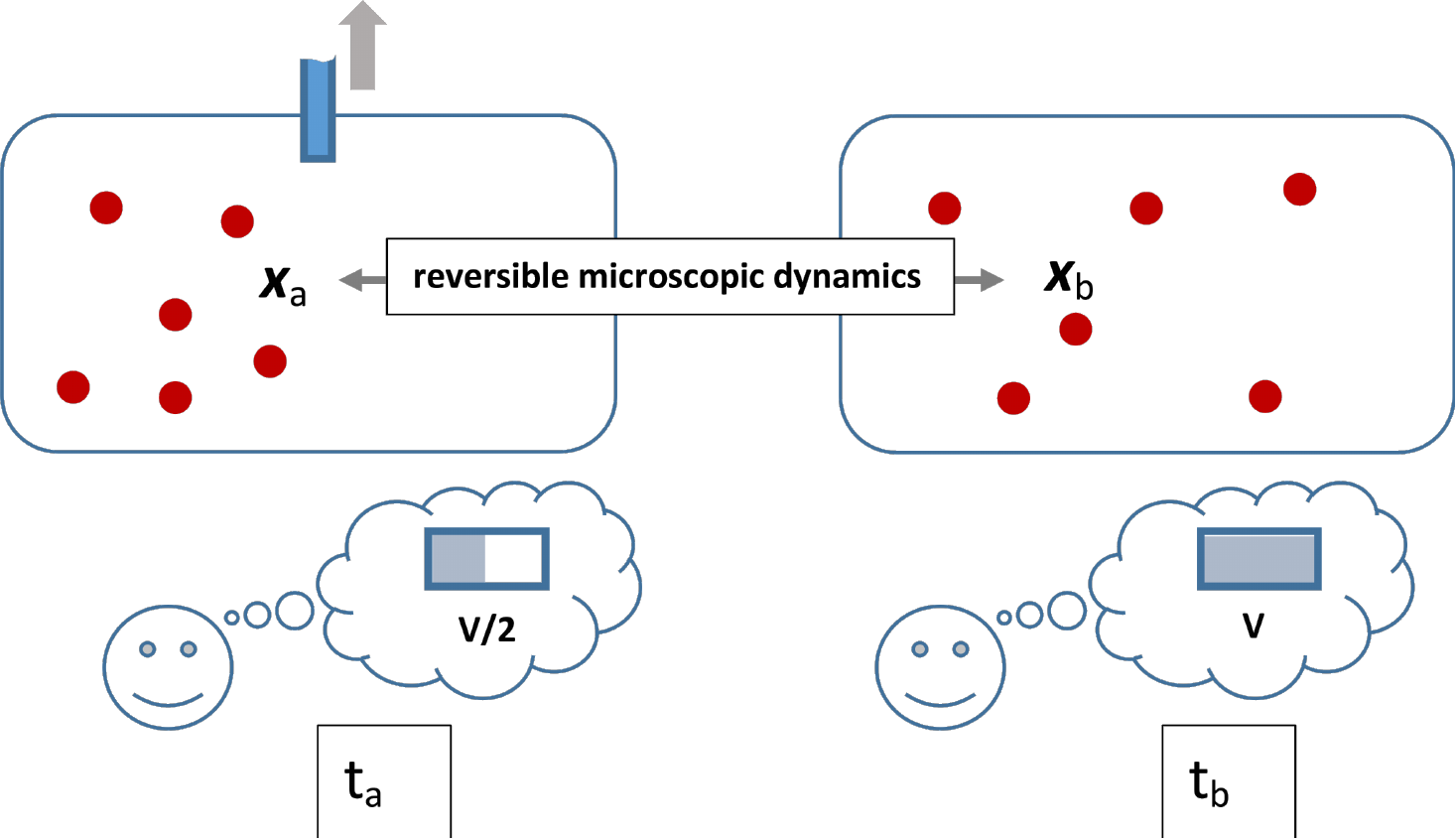}
\caption{Immediately after the removing of the partition the system is closed and the unitary microscopic process, $\xxI\to\xxF$, realizes. At $\tI$, Alice predicts the macroscopic change $V/2\to V$, Bob concludes by the same statistical arguments used at $\tF$ that only the process $V\to V$ was possible.    }
\label{fig:1}
\end{figure}

The closed macroscopic system evolves so that its entropy cannot decrease. It has a convincing statistical explanation  given by L.~Boltzmann in the late 19th century \cite{Boltzmann1896,Lebowitz1993,GLTZ2019,Price2010}. The problem is, however, that the time symmetry of 
laws of microscopic physics guarantees that the same statistical analysis used in the opposite time direction leads to the conclusion that the entropy cannot be lower in the past too \cite{Lebowitz1993,Callender2021}. It can be illustrated on a simple example (see Fig.~\ref{fig:1}). 

Consider two observers detecting \emph{independently} the macroscopic state of a gas in box at two different times. At $\tI$, Alice detects the gas at the left part occupying the volume $V/2$, Bob finds it as filling the whole volume $V$ at $\tF$. Bob does not know the past of the gas but using  statistical analysis he deduces  that the probability that its actual microstate $\xxF$ is a final microstate of a process $V'\to V$, where $V'$ is smaller  then $V$ (e.g. $V'=V/2$), is in  order of $10^{-10^{20}}$, i.e. it is practically impossible.  

 When studying this situation  in a broader  context we see that, before $\tI$, there must exist  some operations guaranteeing the initial conditions of the  gas at $\tI$ \cite{Penrose2005}. Such considerations  lead into the so called \emph{past hypothesis} \cite{Albert2000}  that the past of the system tends to states with an extremely low entropy  \cite{Lebowitz1993,Earman2006}. Using  this hypothesis Bob \emph{knows}  that the past states have lower entropy and cannot derive the previous  (incorrect) result.

Namely any statistical analysis depends on attainable \emph{information}.  The past hypothesis brings general information that prevents us from analyzing the past situation in the same way as the future one. 
Hence  the observer must know beforehand in which time direction the past is. We humans know it though it is not easy to explain why \cite{Szelag2004,Wackermann2005,WittWass2009,MlodBrun2013,CollellFauquet2015}.      In the objective physical description of reality, however, the situation depicted at  
Fig.~\ref{fig:1} should be described without human observers -- Alice and Bob could be robots (computers endowed with some sensors).  What does then mean the conflict of Alice's observation at $\tI$ and Bob's statistical prediction concerning $\tI$?

In this Letter, we analyze such situations from a pure information viewpoint \cite{Jaynes1957,MarNorVed,SagUed2012,DefJar2013,BarSei2014b,ParHorSag2015}. The observer here is a robot  strictly determined by doing only these operations: it performs a measurement at time $t$, gains  coarse-grained information $\info (t)$ about the state of a system, processes it anyway, and records the result. The question is at \emph{which} time this record can be read. 
The answer is simple: if the observed system is macroscopic and it is isolated between times $t$ and $t'$ (its microstate evolves unitary) the record can be read at $t'$ \emph{only} if its entropy fulfills $S(t')\ge S (t)$. If $S (t')> S(t)$ the \emph{time symmetry is broken}: records written at $t$ can be read at $t'$ while records written at $t'$ cannot be read at $t$. 

It nicely corresponds to the fact that we remember the past but not the future. Moreover, the results may have a relation  to the quantum-mechanical phenomenon studied  in Ref.~\cite{Maccone2009}: the decrease of entropy during the evolution of a system quantum-mechanically entangled with its observer  is  accompanied by erasure of records about the observed process (e.g. in observer's brain).

The contribution is organized as follows. First, the concepts like "observation" and "record" are defined by using the special model of the information gathering and utilizing system (IGUS) \cite{Hartle2005}. Information gathered by the IGUS defines the (information) entropy \cite{Jaynes1957} that allows to derive an information form of the fluctuation theorem \cite{Crooks2000,Wang2002,Sev2008,SeifHafeziJarzynski2021}, Eq.~(\ref{PPr}). In the macroscopic limit, it implies the break of time symmetry in  possibility of reading records of the IGUS. It determines the time arrow in the direction of increasing entropy. The problem of Loschmidt's time reverse \cite{Loschmidt1876} (as a spin-echo experiment \cite{Echo2016}) is then discussed in the light of the    
gained result.

\emph{Observation and records}.  Consider  the evolution of a physical system $\syst$ that is observed by an observer $\OS$. The observer may be a human who passively detects the varying current situation, an experimenter or a robot connected with an experimental device measuring given actual parameters, etc. Whichever the case,  $\OS$ gains immediate information $\info (t)$ about $\syst$  at a time $t$ (a set of coarse-grained data concerning the  observed state of $\syst$).  
 
Information $\info (t)$ is transformed  into a complex set of records that become various physical representations of this information \cite{Landauer1999}  and can serve as a memory \cite{Wolpert1992,Holecek2019}:  changes in human brain cells, photos, experimental data saved at hard disks, etc.  
 In Ref.~\cite{Hartle2005}, the concept of information gathering and utilizing systems (IGUS) is introduced to represent a simplified version of  $\OS$. The IGUS has $n+1$ registers. The actual information $\info (t)$ is stored in the register $P_0$ so that the existing  content of $P_0$ is relocated  into $P_1$,  $P_1\to P_2,\ldots ,P_{n-1}\to P_n$,  and the content of $P_n$ is erased. 
 
 The ordered set of registers $P_i$ determines the time orientation: the content of $P_i$ is the past with respect to that of $P_{i-1}$. We assume, however, that the observer has no predetermined time direction (is time neutral). That is why we introduce a time neutral IGUS that has only the register $P_0$. It stores the actual information $\info (t)$. $P_0$ cannot serve as a memory:  if  information from another time, $\info (t')$, is detected  it is automatically stored in $P_0$ and $\info (t)$ is erased.  
 
\begin{figure}[h]
\includegraphics[width=0.4\textwidth]{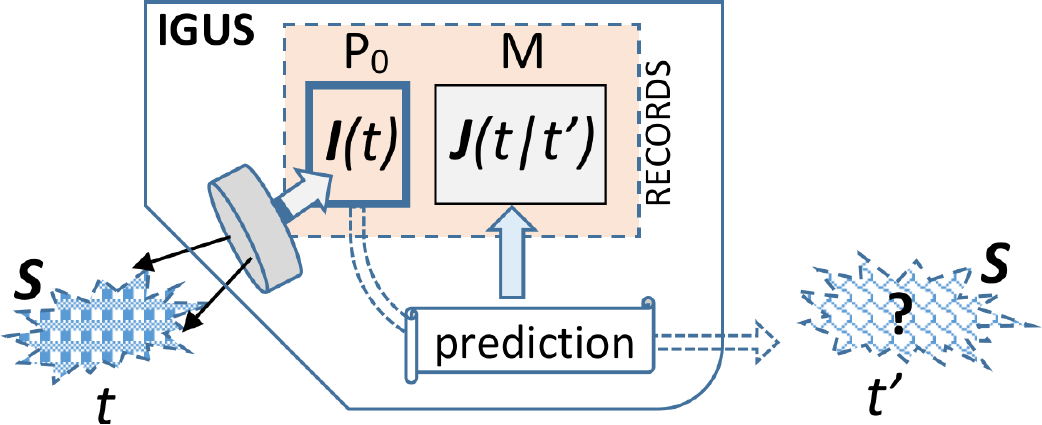}
\caption{Two records in the IGUS "brain": information about the actual situation ($\info (t)$) is in register $P_0$. This record is erased whenever a new observation of $\syst$ is done.  The second register, $\rec$, is a simple memory of IGUS. It can save  information $\info (t_0)$ from an arbitrary time $t_0$ or any transformation (processing) of this information. A typical transformed information is a prediction of possible state of $\syst$ at another time. }
\label{fig:3}
\end{figure}

The memory of IGUS is a single storage $\rec$ in which it can record its  knowledge  about $\syst$. The record exists till new information is recorded into $\rec$. The  knowledge of the IGUS about $\syst$ is  $\info (t)$ (with $t$ being the time of the last observation) or any information transformed from $\info (t)$.   For example, it can recalculate data included in $\info (t)$ by the use of existing physical laws and gain  information $\sel (t|t')$  about  possibly observed state of $\syst$  at another time $t'$ (see Fig.~\ref{fig:3}). We call $\sel (t|t')$ the \emph{prediction} though there is no order of times $t$ and $t'$; obviously $\sel (t|t)=\info (t)$.
 
The state of IGUS is thus defined by the content of $P_0$ and $\rec$, i.e. $\info (t)$ and $\info_{tr}(t_0)$, respectively, where  $\info_{tr}(t_0)$ is a transformed information $\info (t_0)$ (e.g. $\info_{tr}(t_0)=\info (t_0)$).  In the  case when $\info_{tr}(t_0)=\sel (t_0|t)$ the IGUS records two information about  $\syst (t)$. The both information are gained from correct physical observations, possible processed by the use of valid physical laws. It implies that $\info (t)$ and $\sel (t_0|t)$ cannot be in contradiction. 

It is worth emphasizing that the prediction is usually based on incomplete data (it is done only on a coarse-grained knowledge of the system at a given time moment).  Hence its accuracy can be relatively low and it may have sometime only a probabilistic character.  Nevertheless, even so we can imagine  predictions that are clearly inconsistent with the observed data. Excluding such situations (i.e. demand that the IGUS cannot be at a state $[P_0,M]$ in which $P_0$ contradicts  $M$) is thus crucial for connecting a robotic IGUS with a physical reality. Notice that it is also a leading principle in our research  of physical world -- if an inconsistency is found out we must either correct the used laws or look for a mistake in  our evaluation of experiments or observations.

 \emph{Entropy}. 
  From an information theoretic perspective,  entropy is associated with observer's ignorance (lack of information) as to the microstate of the system \cite{Jaynes1957,Crooks1999}. In other words,  the observer has some information about the system, $\info_S$, and entropy is  information that is necessary to "add" to observer's knowledge to determine the actual microstate, $\xx$. 
The entropy thus can be written as $S(\xx |\info_S )$. 
The entropy is a \emph{state} quantity  \cite{Callen1985,LiebYngvason2013} what means that information $\info_S$ must  depend only on actual data gained at a concrete time.  Hence  $\info_S$ is information gathered by our special IGUS, i.e. $\info_S=\info$. 

The concrete form of $\info $ can be very diverse: it can be the actual value of the system microstate, $\xx$,  the value of a thermodynamic quantity $\yw$ (e.g. $\yw =V$), the quantum projector $X =\sum_1^n |\psi_i\rangle\langle\psi_i|$ describing an incomplete knowledge about the system in a mixed state \cite{StrasWint2021}, or parameters of any coarse-graining description \cite{BaronTVEMWG2007}. Whatever the case, we can identify $\info$ with a subset $\Gh (\info)\subset\GA$, where $\GA$ is  the state space of the studied system $\syst$ and  the actual microstate $\xx\in\Gh (\info )$.   
Using the Shannon characterization of information \cite{Shannon1948,LombardiHolikVanni2016} we can identify the entropy of the microstate of the system \cite{Crooks1999},
\begin{equation}\label{infS}
S (\xx| \info) = -k_B\ln p (\xx|\info ),
\end{equation}  
 where $p (\xx |\info )$ is a probability that the system is at the microscopic state $\xx$ if we know that   $\xx\in\Gh (\info)$, and $k_B$ is the Boltzmann constant. Eq.~(\ref{infS}) implies that if $\info_1$ is more precise then $\info_2$, i.e. $\Gh (\info_1)\subset\Gh (\info_2)$, then $S (\xx |\info_1)\le S(\xx |\info_2 )$ (with the use of conditional probabilities).

\emph{Time evolution}.  Imagine an experimental setup arranged in a closed lab with two  observers, Alice and Bob. Alice comes into the lab at $\tI$,  gains information $\infA =\info (\tI)$ about a system $\syst$ and  leaves the lab. Bob does the same at $\tF$ when he  gains  information $\infB =\info (\tF )$ (see Fig.~\ref{fig:1}). There is a single memory cell $\rec$ outside the lab. After leaving the lab, each observer reads $\rec$ and compare its content with own  observation (if it is relevant). Then she/he makes a transformation of her/his observation ($\info \to\info_{tr}$) and records it in $\rec$. 
 The observers are independent and can communicate only via the record in $\rec$. Hence Alice and Bob can be formally identified with one IGUS with a single register $P_0$ including either  "$\infA$ at $\tI$" or   "$\infB$ at $\tF$".

The system $\syst$  is isolated in between  the times when Alice and Bob perform their observations and its evolution is unitary: the microstate  at time $\tI$, $\xxI$, is transformed to the microstate at $\tF$,  $\xxF$, via a one-to-one mapping, $\fun$, defined on  $\GA$, i.e. $\xxF =\fun (\xxI )$. 
Denote $\Gs$ the maximal subset of $\Gh (\infA )$  
so that $\Gf \equiv \fun (\Gs ) \subset \Gh (\infB )$. If $\xxI\in\Gs$ then the conditional probabilities $\pr=p/\pin$ on $\Gs$ and $\pf=p/\pout$ on $\Gf$ fulfill
\begin{equation}\label{prpf}
\pr (\xxI )=\pf (\fun (\xxI )) ,
\end{equation} 
whereas 
$\pin = \sum_{\xx\in\Gs} p (\xx|\infA )$ is the probability that information $ \infA$  gained by Alice at $\tI$ implies that Bob gains information $\infB$ at $\tF$, i.e. $\sel (\tI|\tF ) =\infB$ is valid with the probability $\pin$. Similarly, $\pout =\sum_{\xx'\in\Gs} p (\xx '|\infB )$ is the probability that information $\infB$  gained by Bob at $\tF$ implies that Alice gains information $\infA$ at $\tI$, i.e. $\sel (\tF|\tI )=\infA$ with the probability $\pout$.  
Eqs.~(\ref{infS},\ref{prpf}) relates the probabilities $\pin$ and $\pout$,   
\begin{equation}\label{PPr}
\frac{\pout}{\pin}=\e^{-k_B^{-1}\Delta S},
\end{equation}
where $\Delta S = S(\xxF|\infB )- S (\xxI|\infA )\equiv S(\tF )- S(\tI)$. 

In the context of IGUS it means, for example,  that if the register $P_0$ includes "$\infA$ at $\tI$" it can be written "$\infB$ at $\tF$ with the probability $\pin$" in the  register $M$.  If the system is microscopic or mesoscopic $k_B/|\Delta S|$ is not negligible and the prediction has only stochastic character.

\emph{Macroscopic limit}.    The existence of macroscopic limit means that it is possible to formulate physics in the  limit $k_B\to 0$ so that $\Delta S$ remains nonzero. Hence $|\Delta S|/k_B\to\infty$ which simulates the description of macroscopic processes when $|\Delta S|/k_B \sim 10^{20}$.  In the macroscopic limit, Eq.~(\ref{PPr}) implies that $\pin =0$ whenever  $\Delta S < 0$. Hence  Alice must predict that observed information at $\tF$ must be connected with entropy $S (\tF )\ge S(\tI )$ and  Bob must predict that observed information at $\tI$ must be connected with entropy $S (\tI )\ge S(\tF )$. If $\Delta S\neq 0$ one prediction must be wrong.

  It is exactly the situation depicted at Fig.~\ref{fig:1}: Alice observes (the IGUS has in $P_0$) "$V (\tI) =V/2$". Bob must conclude that the gas cannot occupy the volume less then $V$ at any time, i.e. he records  "$V(\tI )=V$" in $\rec$.  If Alice can read Bob's record (i.e. if there exists the state of IGUS $[P_0,\rec ]=[V(\tI )=V/2,V(\tI )=V]$) we get the situation when the contents of $P_0$ and $\rec$ are in contradiction.
 On the other hand, if Bob reads Alice's prediction  he does not indicate any controversy since  the record in $\rec$  is "$V (\tF )= V'$, $V'\ge V/2$"  and Bob knows  (the IGUS has in $P_0$) that "$V(\tF )=V$".
 
The only conclusion is that the IGUS cannot be in the state 
$[V(\tI )=V/2,V(\tI )=V]$, i.e. 
Alice cannot read Bob's prediction.  It breaks  the  symmetry between times $\tI$ and $\tF$: the record written at $\tI$ can be read at $\tF$ but the record written at $\tF$ cannot be read at $\tI$. This symmetry break has the general validity.  Since information can be transferred  (can \emph{flow}) into a different time only via some records \cite{Maccone2009,Barbour1999} this conclusion means that information concerning macroscopic observations of an isolated system can flow from $t$ to $t'$ only if $S (t ')\ge S(t)$. 
 
This symmetry break  is in  agreement with our everyday observation. Information gained at $t_0$ cannot flow to any time $t < t_0$: we cannot influence the past. On the other hand, sending information to future (via performed records) is a typical human activity: writing, planning,  etc. The message  written  morning can be read the same day afternoon. The opposite order  is impossible.  

It implies why we can remember only the past. Namely any memory is a record in our brain. The record done at $t$ can be read at $t_0$ only if $t < t_0$. If we imagine that Bob is the older Alice (i.e. Alice at $\tF$) then Alice can record the actual situation observed at $\tI$ into $\rec$ (some cells in her brain), i.e. "$V(\tI )=V/2$" in $\rec$. At $\tF$, Alice  gains information $\info (\tF )=V$ (has it in $P_0$). Since $\tF > \tI$ she can read $\rec$, i.e. she \emph{remembers} the situation at $\tI$:  she knows that the gas occupied the volume $V/2$ at $\tI$. 

It seems, however, as a contradiction with Bob's prediction done at $\tF$: $\sel (\tF|\tI) =V$. This prediction, however, is derived (transformed) only from $\info (\tF)$ corresponding to the entropy $S(\xx|\info )$ at $\tF$. But Alice (as older Bob) has information about $\syst$ at $\tF$ that is not $\info (\tF )$ but $\info_\star =(\info (\tF ),\info (\tI ))$. Information $\info_\star$, however, does not define entropy.  Eq.~(\ref{PPr}) then cannot be used (the knowledge of Alice is $\info_\star$ that implies that $\xxI\in\Gs$ and probabilities $\pr$, $\pf$ and Eq.~(\ref{prpf}) must be used instead).

\emph{Time reverse}. Notice that the direction of information flow is given only by the difference of entropy -- no aspects of microscopic dynamics play here a role. A nice illustration is an idea of a hypothetical twin of our universe whose microscopic state is a time reversed state of our universe. The molecules in the box at Fig.~\ref{fig:1} thus have opposite signs of velocites in this universe and the dynamic operator $\fun$ works as follows: $\fun (\xxF\R)=\xxI\R$, where $\xx\R$ is the reverse of the microstate \cite{MaesNetocny2003}. But time does not flow from $\tF$ to $\tI$  since information can flow \emph{only} from $\tI$ to $\tF$ ($S(\tI )< S (\tF)$ is valid in the twin universe too). The passage of time in such a twin universe must be the same as that in ours, i.e. in the agreement with the second law of thermodynamics \cite{Carroll2008}.

\begin{figure}
\includegraphics[width=0.4\textwidth]{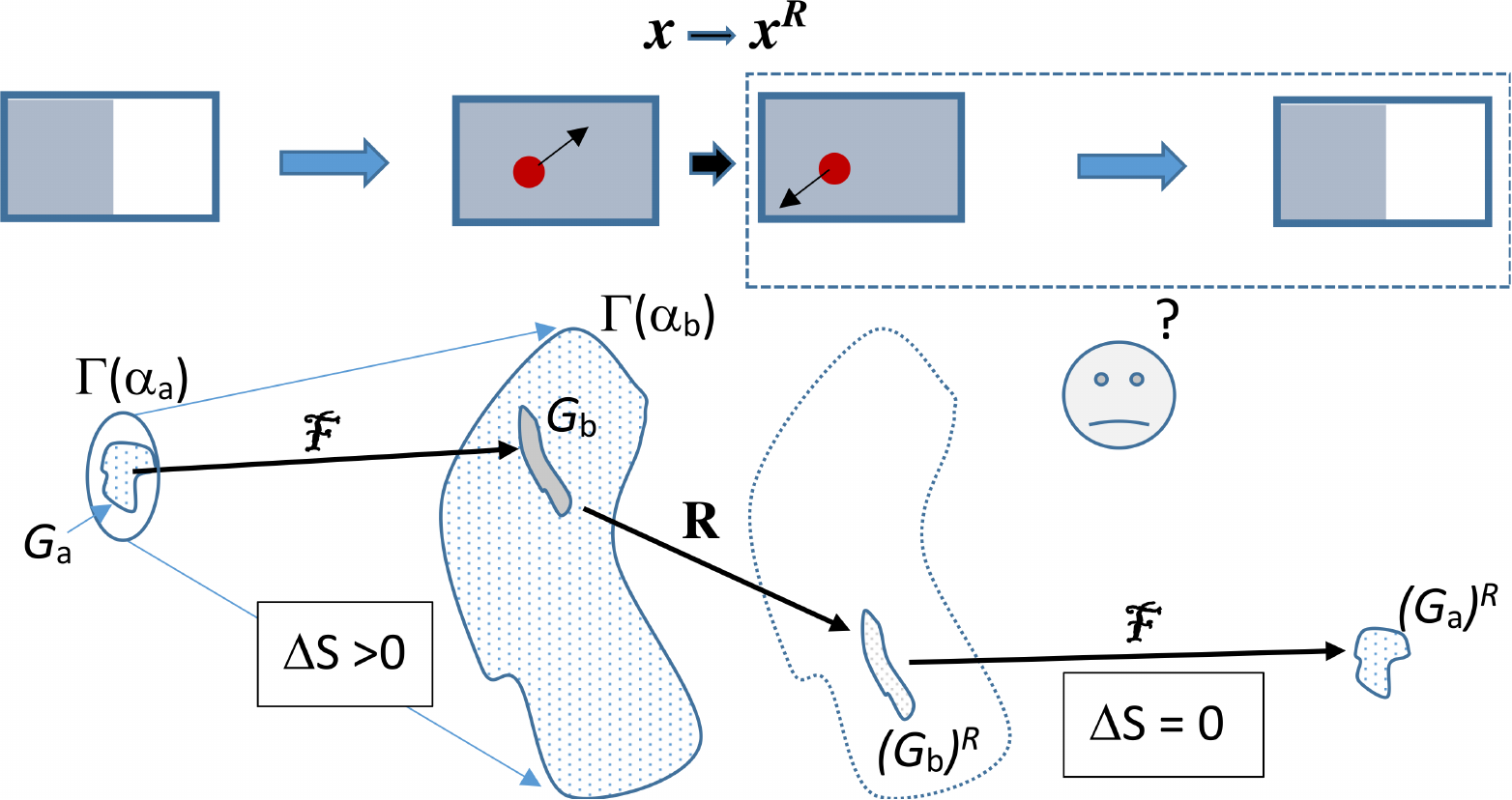}
\caption{The time reversion, $R$, of the microstate at the moment when the gas fills the whole box can bring the gas again to the left part of the box. The process must be strictly unitary: any external perturbation changes a very special initial condition of the second half of the process,  $\xx\R\in(\Gf )\R$, and no backward time evolution can happen. The standard thermodynamic process -- the spontaneous gas expansion -- is not sensitive on the perturbation: nearly any perturbed  microstate tends to $\Gh (\ywF )$. }
\label{fig:2}
\end{figure}

The existence of time reversed states in \emph{our} universe then seems to be paradoxical. Imagine that there exists a possibility of realizing the Loschmidt thought experiment, i.e. performing a sudden change of the sign of velocities of all  molecules, $\xx\to\xx\R$, by an external action with the system at the moment when the gas occupies the whole box after the unitary evolution with the entropy increase, i.e. at time $\tF$ at Fig.~\ref{fig:1}.
The next unitary evolution brings the system into the state $\xxI\R$, i.e. to the macrostate when the gas is in the left half of the box  (see Fig.~\ref{fig:2}).  The passage of time (as information flow) then must have the opposite direction in the stage depicted within the dashed box at Fig.~\ref{fig:2}. How can it be? 

To answer the question we must ask \emph{who} can observe the problematic process, $V\to V/2$. It is important realizing that the initial conditions of \emph{this} process has to be prepared by the standard thermodynamic process $V/2\to V$ and the subsequent  time reversion. Namely the microstate before the reversion has to be within an extremely  small, precisely  given set $\Gf$. 
It implies that the preparing process has to be strictly unitary and no external perturbation can happen. 

After the reversion, the microstate of the system belongs into the small, special set $(\Gf)\R$. The microstates from this set represent the gas occupying the whole volume at the macroscopic scale. But how this macrostate can be observed? It is necessary to break the isolation and interact with the system. This interaction -- whatever small -- perturbs the initial condition, the microstate leaves the set $(\Gf)\R$ and the system does not follow the process $V\to V/2$ (and $V\to V$ is expected instead).

Hence the whole Loschmidt process  must be  realized in an absolute isolation that is not realistic \cite{ECHO}. If so, the observer can still observe only the beginning and the end of the process, i.e. $V/2\to V/2$. It would be a strange situation but it were not in contradiction with the second law of thermodynamics since $\Delta S =0$ during this process. 

\emph{Concluding discussion}.   
The main idea of Boltzmann's statistical program is that the macrostate with a higher entropy is overwhelmingly larger then that with a lower one. Hence the microstate wanders  into this huge set in overwhelmingly many cases \cite{Penrose2005}. There are two problematic points: (i) a certain dependence  on the concrete form of microscopic dynamics \cite{Earman2006}, (ii)  problematic conclusions obtained in the opposite time direction.

The content of Eq.~(\ref{PPr}) is similar: the probability that  $\fun(\xx )$ or $\fun^{-1}(\xx )$ with $\xx$ being a random microstate from a huge subset  $\Gh$ of the state space "hits" a very small target is extremely low. The interpretation is, however, different: $\pin$ or $\pout$ are probabilities of the rightness of \emph{predictions} of an observer who has  information \emph{only} from the time when $\xx\in\Gh$. The result is valid for each unitary dynamics  and  the inconsistency in one time direction has a deep physical meaning in the macroscopic limit. Namely if a consistent physical description of the macroscopic world exists the macroscopic limit cannot give contradictory physical results. 

The assumption that \emph{records} of predictions concerning times with  lower entropy  \emph{cannot be read} at those times gives a consistent picture without contradictions. 
The crucial (and most subtle) point of this consideration is that  Eq.~(\ref{PPr}) is valid if it concerns observers who have information \emph{only} from a single time moment. Pictures like Fig.~\ref{fig:1} are deceptive: they present the situation as if it were known at both times. But information of this kind  cannot be connected with entropy. 
 
 Entropy -- as a state quantity -- must be related to information concerning a concrete time moment. The question is if entropy at $t$ can be defined also from a prediction $\sel (t_0|t)$ that is also information concerning $t$. Let us define 
$S'(t)\equiv S (\xx (t)|\sel (t_0|t))$.  Information  $\sel (t_0|t)$ is, however, only transformation of information $\info (t_0)$ which implies that it cannot be more precise then $\info (t_0)$, i.e.  $S' (t) \ge S (t_0 )$. If 
$S(t )< S(t_0)$, i.e. we do the prediction towards the past, it cannot  be in accord with the observation at $t$ since  $S'(t)$ is \emph{always} different from $S (t)$ ($S (t) <S'(t)$). Towards the future,  $S'(t)$ can equal $S(t)$ since $S(t)\ge S(t_0)$.   

This consideration outlines again the important role of predictions (i.e. processed information gained from external impulses at different times). It is worth stressing that information about a concrete event that happens  at a given time is usualy 'read' by humans via  various physical stimuli  (visual, auditory,  etc.) that are recorded  at different times   \cite{Szelag2004}.  The permanent 'recalculation' of records (in the sense of $\info \to \info_{tr}$ defined above) thus appears as important element of our perception of time. 

Concerning the whole concept we may ask for the role of probabilities since the observer may not have any idea about probabilities of possible microstates. The answer is that Eq.~(\ref{PPr}) is used here only in the macroscopic limit in which a concrete probability distribution does not play a role (entropy $S (\xx|\info )$ can be defined without using the concept of probabilities, the crucial result that $\pin$ or $\pout$ tends to zero is then equivalent to  $\info\to\infty$ \cite{Holecek2021}). 

There are many other questions connected with the derived break of time symmetry, especially the relation to the perceived time passage or the concept of time on a relativistic spacetime block \cite{Barbour1999,Rovelli2004,BLOCK}. It would be also appealing to find an interconnection with the past hypothesis (that can be interpreted as new additional information) including its cosmological context \cite{Price2010,Earman2006,Carroll2008,Wald2006}. We also do not explain here the mechanism why the record 'vanishes' at times when the entropy is lower. It might have an interesting relation to quantum-mechanical phenomenon analyzed in Ref.~\cite{Maccone2009}. Our approach may also contribute to the question of emergence of the macroscopic world \cite{ParHorSag2015}. The information description of the macroscopic limit can  be instructive here.

\emph{Acknowledgment}. 
The author is indebted to J\'{a}n Min\'{a}r for his help and constructive discussions, and to Phill Strasberg for inspirational and critical comments concerning the foundations of thermodynamics and statistical physics. The work is supported by the New Technologies Research Center of the West Bohemia University in Pilsen.


\begin{thebibliography}{999}
\bibitem{BLOCK}  The IGUS on the spacetime block can be described as a set of couples $[P_0(\tau ),\rec (\tau)]$ where $\tau$ is the parameter of individual spacelike slices $\Sigma (\tau)$. Our main result then can be  formulated so that $\rec (\tau)$ cannot record information from slices $\Sigma (\tau ')$ with $\tau '> \tau$ if the slices are ordered so that the entropy of an isolated system does not decrease if $\tau$ is increasing.
\end{thebibliography}
\end{document}